\documentclass[twocolumn,showpacs,showkeys,preprintnumbers,pre,amsmath,amssymb,superscriptaddress,longbibliography]{revtex4-1}
\usepackage[colorlinks=true,linkcolor=blue,urlcolor=blue,citecolor=blue]{hyperref}
\usepackage{graphicx}
\usepackage{natbib}
\usepackage{color}
\usepackage{amsmath}
\usepackage{enumerate}
\usepackage{hyperref}
\usepackage{lipsum}
\usepackage[toc,page]{appendix}
\usepackage[normalem]{ulem}
\def\strutdepth{\dp\strutbox}
\def\nw#1{\strut\vadjust{\kern-\strutdepth\vtop to0pt{\vss\hbox to\hsize
{\hskip\hsize\hskip5pt$\leftarrow$\hss\strut}}}{\em #1}}

\newcommand{\tEff}{\tau_{\mbox{\scriptsize{eff}}}}
\newcommand{\TFict}{T_{\mbox{\scriptsize{F}}}}

\newcommand{\peq}{p_{eq}}

\newcommand{\Df}{d(\vec{x},t)}

\newcommand{\del}[1]{}  

\begin{document}

\title{Spatial heterogeneities in structural temperature cause Kovacs' expansion gap paradox in aging of glasses}

\author{Matteo Lulli}
\email{matteo.lulli@gmail.com}
\affiliation{Department of Applied Physics, Hong Kong Polytechnic University, Hong Kong, China}

\author{Chun-Shing Lee}
\affiliation{Department of Applied Physics, Hong Kong Polytechnic University, Hong Kong, China}

\author{Hai-Yao Deng}
\affiliation{School of Physics and Astronomy, The University of Manchester, Manchester M13 9PL, United Kingdom}
\author{Cho-Tung Yip}
\affiliation{School of Science, Harbin Institute of Technology, Shenzhen Graduate School, Shenzhen, Guangdong 518055, China}
\author{Chi-Hang Lam}
\affiliation{Department of Applied Physics, Hong Kong Polytechnic University, Hong Kong, China}

\date{\today}

\begin{abstract}
  Volume and enthalpy relaxation of glasses after a sudden temperature change has been extensively studied since Kovacs’ seminal work. One observes an asymmetric approach to equilibrium upon cooling versus heating and, more counter-intuitively, the expansion gap paradox, i.e. a dependence on the initial temperature of the effective relaxation time even close to equilibrium when heating. Here we show that a distinguishable-particles lattice model can capture both the asymmetry and the expansion gap. We quantitatively characterize the energetic states of the particles configurations using a physical realization of the fictive temperature called the structural temperature, which, in the heating case, displays a strong spatial heterogeneity.  The system relaxes by nucleation and expansion of warmer mobile domains having attained the final temperature, against cooler immobile domains maintained at the initial temperature. A small population of these cooler regions persists close to equilibrium, thus explaining the paradox.
\end{abstract}

\maketitle
Kovacs' series of experiments~\cite{Kovacs1964} is fundamental to our present understanding of aging and memory properties in glassy materials~\cite{Roth2016,Angell2000Rev,Hodge1994}. In~\cite{Kovacs1964}, the volume relaxation of polymer glasses has been analyzed by performing rapid temperature changes, or \emph{temperature jumps}, focusing on experimental protocols implementing one or two successive temperature shifts. The renowned Kovacs effect, observed in experiments involving two successive rapid temperature changes, or a double temperature jump, has been theoretically studied using empirical mean-field models including the Tool-Narayanaswamy-Moynihan (TNM)~\cite{Tool1946, Narayanaswamy1971, Moynihan1976} and the Kovacs-Aklonis-Hutchinson-Ramos (KAHR)~\cite{KAHR1979} models. A temperature jump more precisely acts directly only on the phonon temperature. The Kovacs effect can be satisfactorily accounted for in these theories using a fictive temperature $\TFict(t)$, which describes some internal state of the material with a dynamics generally lagging behind that of the phonons~\cite{Tool1946}. In contrast, the \emph{expansion gap paradox}, also called the $\tEff$ \emph{paradox}, from single-jump experiments is much more puzzling~\cite{McKenna1995, McKenna1999, Kolla2005, Hecksher2010, Hecksher2015, Banik2018, Struik1997_1, Struik1997_2}. Specifically, the effective relaxation rate $\tEff$ of the polymeric system studied by Kovacs, after the temperature jump, depends persistently on the initial temperature, and apparently, even arbitrarily close to equilibrium. Such a strong material memory, however, cannot be reproduced by TNM or KAHR models, and has only been accounted for by their stochastic counterparts, namely the stochastic version of a free-volume model~\cite{Robertson1984} and, more recently, the stochastic constitutive model (SCM)~\cite{Medvedev2015}. The reasons for the failure of simple mean-field models and why stochastic fluctuations are extraordinarily important for this experimental protocol are not well understood.

Here, we successfully reproduce Kovac's expansion gap for the first time using a microscopic particle model, going beyond mean-field descriptions. Specifically, we adopt the distinguishable-particle lattice model (DPLM)~\cite{DPLM2017}. The phonon temperature, which is subjected to a single jump, is modeled by the bath temperature of the kinetic Monte Carlo simulation algorithm of the DPLM. We observe an expansion gap in the system energy relaxation, analogous to enthalpy relaxation in experiments~\cite{Montserrat1994}. By studying spatial profiles of particle displacements and interactions, we provide an intuitive resolution of the paradox. 

The recently introduced DPLM displays many features of particle dynamics characteristic of glasses and possesses exactly solvable equilibrium statistics~\cite{DPLM2017}. Moreover, a wide range of values of the fragility index can be obtained by varying the interaction energy distribution of the model~\cite{lee2019}. In the simulations we have performed, the DPLM is defined on a $2$-dimensional square lattice of linear size $L = 100$. The sites are occupied by $N$ distinguishable particles, each of them associated to a \emph{unique label} ranging from 1 to $N$. Then, $s_i =1,\ldots,N$ denotes the particle at site $i$. The key feature of the model is that each particle is coupled to its nearest neighbors by means of \emph{site-} and \emph{particle-dependent} random interactions: a four-indices interaction energy $V_{ij s_i s_j}$ is associated to the particles $s_i$ and $s_j$ sitting at sites $i$ and $j$. In order to simulate the hopping dynamics of the particles we allow for the presence of empty sites or \emph{voids}. Considering a void density $\phi_v = 0.005$, we allow the presence of $N_v = 50$ voids with default label $s_i = 0$ so that $L^2=N+N_v$. One can write the system energy as
\begin{equation}
  \label{eq:E}
  E = \sum_{\langle ij \rangle'} V_{ij s_i s_j},
\end{equation}
where the sum $\sum_{\langle ij \rangle'}$ is restricted to occupied nearest neighboring sites.  The interactions are symmetric under concurrent exchange of spatial and particle indices, i.e. $V_{ijkl} = V_{jilk}$. The entire set of possible interactions $\{V_{ijkl}\}$ is drawn according to an \emph{a priori} probability distribution $g(V)$ and it is \emph{quenched}. For each particle configuration, there is a set $\{V_{ij s_i s_j}\}$ of interactions which are referred to as \emph{realized}. It has been proved in~\cite{DPLM2017} that at equilibrium the probability distribution of the interactions factorizes over all occupied neighboring sites. The equilibrium probability distribution of any realized interaction at temperature $T$ follows the form
\begin{equation}
  \label{eq:peq}
  \peq(V,T) = \frac{1}{\mathcal{N}(T)}\,g(V)\,e^{-V/k_BT}
\end{equation}
where $\mathcal{N}(T)$ is a normalization constant. For the rest of the paper, we will work in natural units with $k_B = 1$. As in~\cite{DPLM2017}, we take $g$ as a \emph{uniform} distribution defined on the interval $V \in [V_0,V_1]$, with $V_1=-V_0=0.5$. In this case, the equilibrium distribution in Eq.~\eqref{eq:peq} yields a simple exponential dependence on the interaction energy. The simulations are performed by means of a kinetic Monte Carlo algorithm for activated hopping dynamics in which each particle can hop to the position of a neighboring void with a rate
\begin{equation}
  \label{w}
 w = w_0 \exp\left[-\frac{1}{T}\left(E_0 + \frac{\Delta E}{2}\right)\right] ,
\end{equation}
where $\Delta E$ is the energy change of the system induced by the hop. We set $w_0 = 10^6$ and $E_0 = 1.5$ so that $E_0 + \Delta E/2 \geq 0$.

\begin{figure}[!ht]
  \includegraphics[scale=0.75]{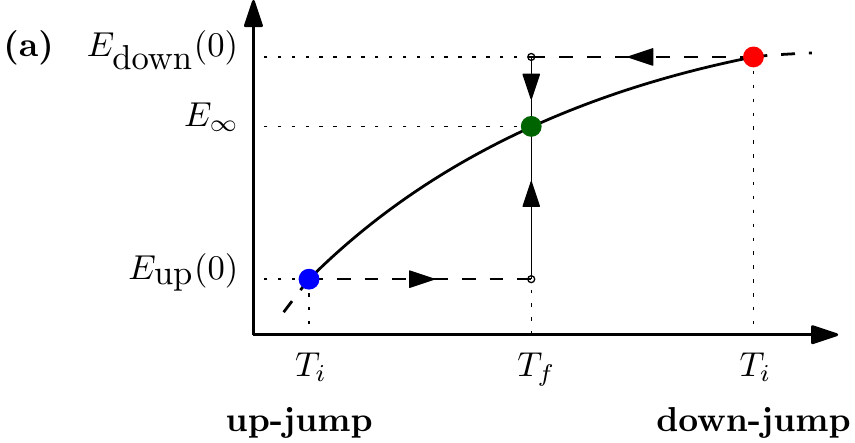}
  \includegraphics[scale=0.6]{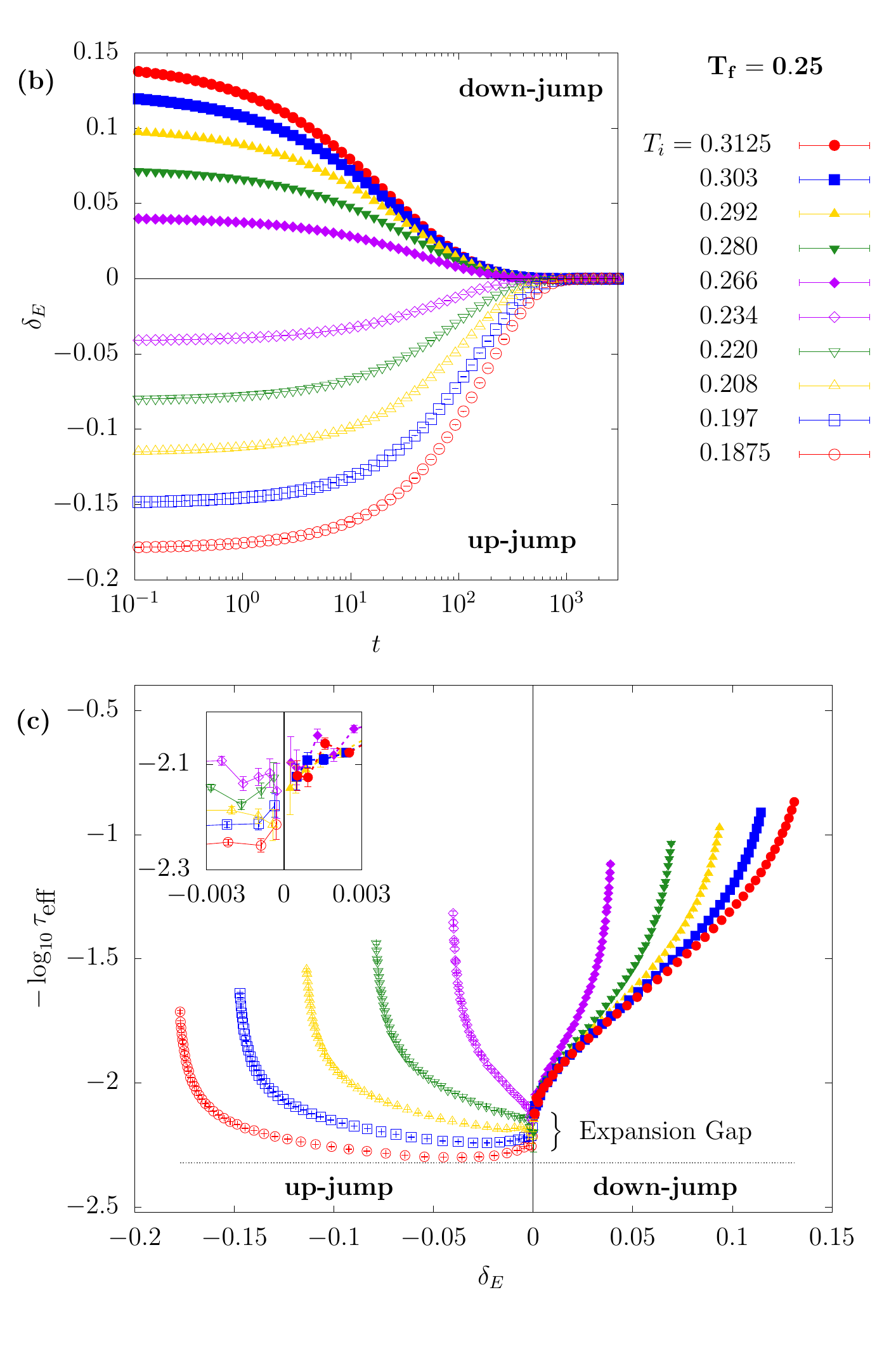}
  \caption{Panel (a): Schematic diagram for single-temperature jump protocol for equilibrium samples at initial temperature $T_{i}$ which are then cooled (for $T_i>T_f$, i.e. down-jump) or heated (for $T_i<T_f$, i.e. up-jump) to the final temperature $T_f$. Panel (b): data from  DPLM simulations for $T_f = 0.25$ and different values of $T_i$. The asymmetry of the approach between up- and down-jumps is observed. Panel (c): results on $\tEff$ measured using data in middle panel. Data close to equilibrium with $|\delta_E(t)| \le 0.003$ are shown in the inset.}\label{fig:gap}
\end{figure}
\begin{figure*}[!ht]
  \begin{center}
    \includegraphics[scale=4.8]{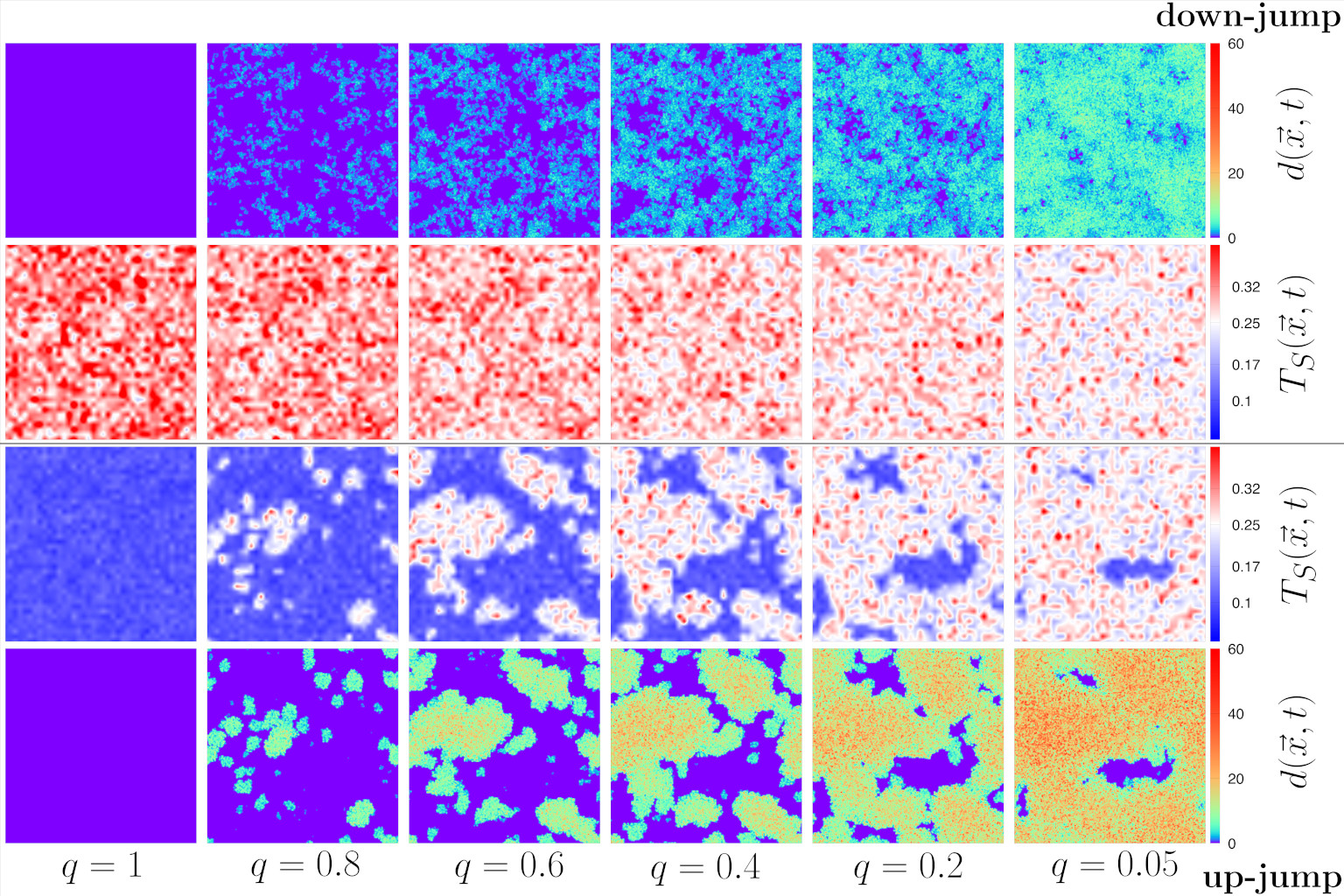}
  \end{center}
  \caption{Snapshots of structural temperature $T_S(\vec{x}, t)$ (2nd and 3rd rows) and  particle displacement $\Df$ (1st and 4th rows) for down-jump from $T_i=0.3125$ (1st and 2nd rows) and up-jump from $T_i=0.1$ (3rd and 4th rows) to final temperature $T_f = 0.25$ in a system of linear size $L=200$ with coarse graining scale $\ell=5$. Different columns refer to average overlap $q = 1, 0.8, 0.6, 0.4, 0.2$ and 0.05 (from left to right) corresponding to increasing  time $t$. Both $T_S(\vec{x}, t)$ and $\Df$ show relatively homogeneous evolution for the down-jump, but strongly heterogeneous evolution with large domains for the up-jump. Finally, one can notice for the up-jump case a clear coincidence between immobile domains and low structural temperature domains.}\label{fig:fields}
\end{figure*} 

In our simulations, we implement the single-jump protocol analyzed in~\cite{Kovacs1964} and displayed in Fig.~\ref{fig:gap}(a) starting from equilibrium configurations~\cite{DPLM2017} at some initial temperatures $T_i$. Then, the bath temperature $T$ in Eq.~\eqref{w}, representing the phonon temperature, is set instantaneously at the final temperature $T_f$. It is a common practice to refer to the cases $T_i > T_f$ (cooling protocol) as \emph{down-jumps}, and to the cases $T_i < T_f$ (heating protocol) as \emph{up-jumps}. 

Analogously to~\cite{Kovacs1964}, we study the fractional deviation $\delta_E(t)$ of the system energy $E(t)$ from its equilibrium value $E_\infty$ at  $T_f$ given by 
\begin{equation}\label{eq:delta}
  \delta_E(t) = \frac{E(t) - E_\infty}{|E_\infty|},
\end{equation}
where $E(t)$ and $E_\infty<0$ are computed using Eq.~\eqref{eq:E} and are averaged over $2^{17} \simeq 1.3 \times 10^{5}$ independent runs with different random number seeds. The analytical expression for $E_\infty$ is reported in Eq.(2) of the Supplemental Material and shows agreement with the numerically computed values. Fig.~\ref{fig:gap}(b) reports the time evolution of $\delta_E(t)$ for a set of \emph{symmetric} temperature jumps, which are chosen to be of similar relative magnitudes as in~\cite{Kovacs1964}. As shown in Fig~\ref{fig:gap}, $\delta_E$ from DPLM simulations closely resembles experimental results in~\cite{Kovacs1964} and, in particular, correctly reproduces the \emph{up-down asymmetry} of the approach to equilibrium. The initial asymmetry at time 0 is simply due to an equilibrium heat capacity decreasing with $T$, of which we report the analytical expression in Eq.(3) of the Supplemental Material. However, the much slower relaxation for the up-jumps compared with the down-jumps, successfully reproduced here, is non-trivial and has been the focus of many studies~\cite{McKenna1995, McKenna1999, Kolla2005, Hecksher2010, Hecksher2015, Banik2018, Struik1997_1, Struik1997_2}. 

We next define the \emph{effective relaxation time} $\tEff$ as in~\cite{Kovacs1964}
\begin{equation}\label{eq:t_eff}
  \tEff^{-1}(t) = -\frac{1}{\delta_E(t)}\frac{\mbox{d}\delta_E(t)}{\mbox{d} t},
\end{equation}
which would reduce to a constant for an exponentially decaying $\delta_E(t)$. Averages and errors for $\tEff$ have been computed with the jackknife resampling method. Results for $\tEff(t)$ against $\delta_E(t)$ are reported in Fig.~\ref{fig:gap}(c) and they show very similar features to those reported in~\cite{Kovacs1964}.
Most importantly, we observe as in~\cite{Kovacs1964} that the data for $\tEff$ have not converged to a single limiting value independent of $T_i$, even very close to equilibrium at $|\delta_E| \simeq 0$, creating the \emph{expansion gap paradox}. The inset reports the data close to equilibrium: while the down-jump data show a clear convergence among themselves and with respect to the up-jump data at small jumps, convergence is not observed in the large up-jump cases even for the smallest $\delta_E$ studied. To the best of our knowledge, among the constitutive models~\cite{Tool1946, Narayanaswamy1971, Moynihan1976,KAHR1979,Medvedev2015,Robertson1984} only the stochastic free-volume model~\cite{Robertson1984} and the SCM~\cite{Medvedev2015}, accounting for dynamic heterogeneities, have been able to reproduce the gap. Being able to qualitatively recover the most important experimental features by means of a microscopic particle model is clearly important for a deeper understanding of the aging dynamics. Nevertheless, $\tEff^{-1}$ in Fig~\ref{fig:gap}(c) does not exhibit a gentle rise at intermediate values of $\delta_E$ for the up-jumps as observed in experiments \cite{Kovacs1964}. This may happen because we have adopted, for simplicity, a constant void density $\phi_v$ in our simulations, which should instead increase upon heating. Such an increase of $\phi_v$ would hence yield a faster dynamics.

Compared to the constitutive models, an advantage of our approach is that it allows us to analyze the differences between up- and down-jump dynamics from the \emph{real-space} perspective, going beyond mean-field descriptions. We define a \emph{local particle displacement} $\Df = |\vec{x}-\vec{x}_0|$ as the distance of a particle located at $\vec{x}$ at time $t$ relative to its position $\vec{x}_0$ at time $0$ at which the temperature jump is imposed. If $\vec{x}$ is vacant at time $t$, we put $\Df = 0$ for simplicity. It is also useful to define a local particle persistence, i.e. an \emph{overlap} field, $\tilde q(\vec{x}, t)$ as
\begin{equation}
  \tilde q(\vec{x}, t) = \left \{
  \begin{array}{@{}rl@{}}
    1 & \mbox{if}\quad d(\vec{x},t) = 0\\
    0 & \mbox{if}\quad d(\vec{x},t) > 0
  \end{array}
  \right.
\end{equation}
\begin{figure}[!ht]
  \hspace*{-0.5cm}
  \includegraphics[scale=0.6]{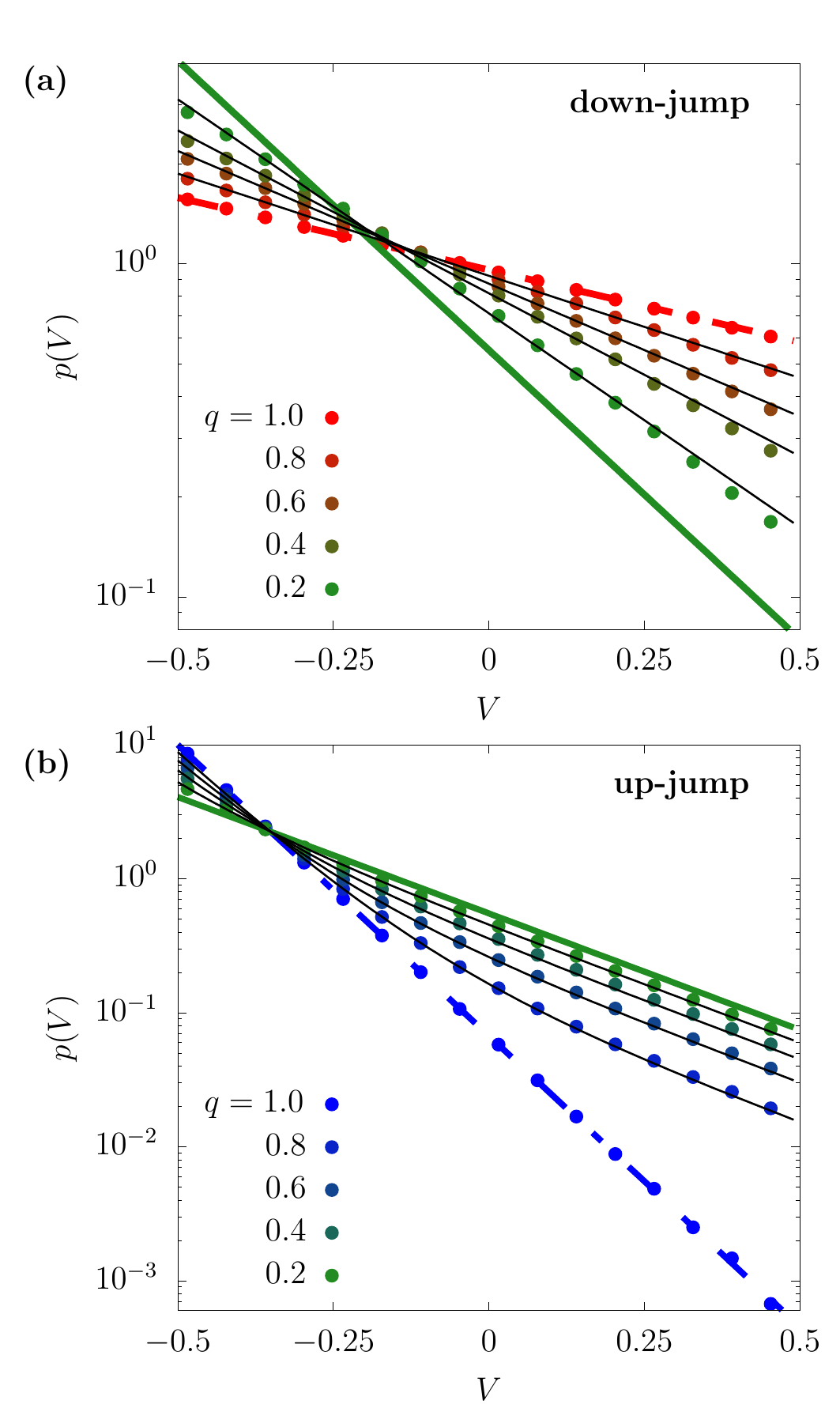}
  \caption{Panels (a) and (b): semi-log plot of the $p(V)$, in colored dots, for five values of the overlap $q=1,0.8,0.6,0.4,0.2$, for the down-jump protocol with $T_i = 1.0$ and the up-jump one with $T_i = 0.1$ to the common final temperature $T_f = 0.25$. The final equilibrium distribution is drawn in green solid lines while the initial distributions are drawn in dashed red and blue lines. Panel (a): Down-jump data superposed to single-temperature fits in black lines, showing a good agreement. Panel (b): Up-jump data superposed to Eq.~\eqref{dualexponential}.}\label{fig:histos}
\end{figure}
The average overlap $q(t)$, which equals $\tilde q(\vec{x},t)$ averaged over sites occupied at $t$, gives the fraction of particles still located at their original positions at time $t$. In Fig.1 in the Supplemental Material we show $q(t)$ against $t$.

In Fig.~\ref{fig:fields} (and in the supplementary videos \emph{supvideo\_up.mp4} and \emph{supvideo\_down.mp4} together with the voids positions), we report the evolution of the local displacement $d(\vec{x},t)$ for selected values of the average overlap $q$  which provides a useful measure of the progress of the relaxation. Large jump magnitudes are used so that qualitative features of the dynamics can be more clearly observable. Fig.~\ref{fig:fields} shows that the growth of $d(\vec{x},t)$ is much more heterogeneous in the up-jump case. Well-separated domains with highly mobile particles nucleate and invade other immobile domains. Hence, a strong spatial heterogeneity dominates the up-jump relaxation. 

In order to understand the emergence of the significantly more heterogeneous up-jump dynamics, we study the energy states of the particle configurations by analyzing the probability distribution $p(V)$ of the realized interactions $V_{ij s_i s_j}$. Results are reported in Fig.~\ref{fig:histos}. In the panels (a) and (b), the computed $p(V)$, for up- and down-jump dynamics respectively, are reported for different values of the average overlap $q$. The final equilibrium distribution $\peq(V,T_f)$ given in Eq.\eqref{eq:peq} is also reported, while the initial ones, i.e. $\peq(V,T_i)$ are reported in \emph{red} and \emph{blue} dashed lines for down- and up-jump respectively. Fig.~\ref{fig:histos}(a) shows that in the down-jump case the evolution of the probability distribution occurs simultaneously for the whole range of $V$. Furthermore, $p(V)$ can be well approximated by a single equilibrium distribution $\peq(V,T)$, with $T$ decreasing monotonically with $1 - q$ and thus also with $t$,  yielding reasonable fits. In contrast, for the up-jump data reported in Fig.~\ref{fig:histos}(b), there is a remarkable difference between the evolution of high- and low-energy interactions: For example for $q = 0.8$, the distribution $p(V)$ at $V \gtrsim 0.25$ has already attained the same slope in the semi-log plot as the equilibrium one at the final temperature $T_f$, while values of $p(V)$ for low-energy interactions $V \alt -0.25$ are still very close to those of the initial temperature $T_i$. Indeed, $p(V)$ for a wide range of $q$ can be very well fitted by a superposition of two equilibrium distributions 
\begin{equation}
\label{dualexponential}
p(V) = q\cdot \peq(V,T_i) + (1-q)\cdot \peq(V,T_f).
\end{equation}
This result suggests that the particle configurations in the \emph{mobile regions}, whose relative extent is $1-q$, have reached equilibrium at the \emph{final temperature} $T_f$ while the \emph{immobile regions}, whose extent is $q$, have an interaction population distributed according to $\peq$ at the \emph{initial temperature} $T_i$.

The previous results suggest the existence of a strong spatial heterogeneity in the distribution of the realized particle pair-interactions only for the up-jump case. As a further step, we compute from the interactions a temperature $T_S$ we call the \emph{structural temperature}. In physical terms, $T_S$ measures how well particles are locally packed, and hence a low temperature, for example, corresponds to a better bonded and more stable configuration. Such a definition can be applied to a wide range of materials and it should not be regarded as specific to the present case. For the DPLM, we define it as a local temperature $T_S(\vec{x}, t)$ at position $\vec{x}$ based on the interaction $\overline{V}_\ell(\vec{x},t)$ averaged over a square domain of linear size $\ell$ centered at $\vec{x}$. Requiring that $T_S$ should coincide with the bath temperature within statistical fluctuations at equilibrium, we define $T_S$ and solve for it numerically from $\overline{V}_\ell(\vec{x},t) = \int\mbox{d}V\,V\,\peq(V,T_S)$. Note that $T_S$ is analogous to Tool's fictive temperature~\cite{Tool1946}, local values of which have also been studied before~\cite{Keys2013,Wisitsorasak2014}. 

Figure~\ref{fig:fields} also shows (along with the supplementary videos \emph{supvideo\_up.mp4} and \emph{supvideo\_down.mp4}) the evolution of $T_S(\vec{x}, t)$ for ${\ell=5}$. By a direct comparison with the local displacement $d(\vec{x},t)$, we see that the evolution of $T_S(\vec{x}, t)$ is spatially homogeneous for the down-jump. For the up-jump,  high-$T_S$ domains with $T_S \simeq T_f$ develop in good spatial correspondence with the highly mobility regions. The immobile regions in contrast maintain the initial temperature, i.e. $T_S \simeq T_i$. These results are fully consistent with the good fits to $p(V)$ in Fig.~\ref{fig:histos}(b) using Eq.\eqref{dualexponential}. Moreover, some low-$T_S$ immobile domains remain even at the very late stage of relaxation at $q=0.05$, thus constituting a remnant of the initial temperature acting on the dynamics, although equilibrium is already reached almost everywhere else. 

The structural temperature heterogeneity observed for the up-jumps can be understood in terms of a stability argument of propagating fronts as follows. First, the heating up of a glass is an auto-catalytic process, since the excitation of particle arrangements to higher-energy configurations speeds up the particle dynamics and hence provides a positive feedback to the further warming of the system. In $d$ dimensions, $T_S(\vec{x}, t)$ can be seen as a succession of equal-time $d$-dimensional surfaces in a $(d+1)$-dimensional space, representing a front propagating upwards from $T_i$ to $T_f$. The propagation is driven by the energy influx from the bath and is stochastic because of the intrinsic noise of the particles dynamics. The evolution of the surfaces is unstable against small perturbations, meaning that a locally out-stretched (warmer) region will further advance much faster towards the final value $T_f$ as the auto-catalytic nature of the dynamics amplifies the perturbations. For very low $T_i$, implying an extreme sensitivity of the dynamics on temperature, $T_S$ can comparatively quickly reach $T_f$ in localized domains, while being practically stuck at the initial value $T_i$ elsewhere. This explains the nucleation of $T_f$ domains in a background of $T_i$ regions. The fast dynamics in $T_f$ domains enhances the heating-up of neighboring regions, inducing domain-wall motions. Due to the very stable configurations of the $T_i$ regions, the domain invasion can be a slow process compared with the relaxation dynamics in the $T_f$ domains. Therefore, the particle displacement $d(\vec{x},t)$ can become very large in the mobile $T_f$ domains even close to their domain boundaries as observable in Fig.~\ref{fig:fields}. By contrast, cooling for the down-jump protocol is instead an auto-retarding process so that the downward propagating front $T_S(\vec{x}, t)$ is stable against perturbations. The dynamics is thus overall homogeneous with relatively uniform $T_S(\vec{x}, t)$ as shown in Fig.~\ref{fig:fields}.

In summary, Kovacs' expansion gap paradox in energy relaxation is reproduced based on kinetic Monte Carlo simulations of a particle model in two-dimensions. 
A structural temperature is introduced to characterize the energy states of the particle configurations. After an up-jump of the bath temperature from $T_i$ to $T_f$, a large spatial heterogeneity is observed in both local particle displacement and local structural temperature. The evolution of the latter is characterized by the nucleation and coarsening of $T_f$ domains invading the original $T_i$ domains. Relaxation dynamics persistently depend on $T_i$ because isolated  $T_i$ domains survive even close to the end of the relaxation. This leads to strong memory effects and explains the paradox. We argue that the strong spatial fluctuations are caused by a spatial instability due to the auto-catalytic nature of the heating of glasses. In contrast, for a temperature down jump, the auto-retarding nature of cooling leads to a stable and thus homogeneous evolution of the structural temperature, resulting in weak memory effects and converging relaxation rates. 

The structural temperature introduced in this work is, to the best of our knowledge, the first example of a physical realization of Tool's fictive temperature~\cite{Tool1946}. As an advancement, it is measurable from particle simulations based on well-defined microscopic dynamics, in contrast to the fictive temperature which follows separate empirical evolution rules \cite{Hodge1994}. Our results stress the importance of its spatial heterogeneity in understanding the expansion gap paradox. This naturally explains  why mean-field models with a global fictive temperature~\cite{Tool1946, Narayanaswamy1971, Moynihan1976,KAHR1979} in general have difficulty reproducing the paradox. It also justifies the results of the stochastic models~\cite{Robertson1984,Medvedev2015} in which different stochastic realizations empirically represent local regions at different stages of evolution. We believe that the structural temperature will be of general importance in the study of non-equilibrium behaviors of glasses. It should be of interest to measure it experimentally by means of, for example, electron correlation microscopy~\cite{Zhang2018}.

Kovacs' experiments are important because exceptional material properties often provide the deepest insights. 
Overcoming the long-standing challenge of reproducing the expansion gap using a microscopic particle model, our results do not only provide a possible intuitive understanding of the paradox, but also support the validity of the DPLM as a reliable computational tool for studying glassy dynamics. 
Finally, our findings should also be useful to scrutinize and to further develop theoretical approaches on glasses~\cite{berthier2011review,garrahan2011review}.

\section{Acknowledgements}
We gratefully acknowledge Chor-Hoi Chan, Haihui Ruan and Giorgio Parisi for interesting discussion and comments. We thank the support of Hong Kong GRF (Grant 15330516) and PolyU (Grants 1-ZVGH and G-UAF7).

\bibliography{references}

\end{document}